# Solar Power Prediction Using Satellite Data in Different Parts of Nepal

Raj Krishna Nepal, Bibek Khanal, Vibek Ghimire, Kismat Neupane, Atul Pokharel, Kshitij Niraula, Baburam Tiwari, Nawaraj Bhattarai, Khem N. Poudyal, Nawaraj Karki, Mohan B Dangi, John Biden

**Abstract**

Due to the unavailability of solar irradiance data for many potential sites of Nepal, the paper proposes predicting solar irradiance based on alternative meteorological parameters. The study focuses on five distinct regions in Nepal and utilizes a dataset spanning almost ten years, obtained from CERES SYN1deg and MERRA-2. Machine learning models such as Random Forest, XGBoost, K-Nearest Neighbors, and deep learning models like LSTM and ANN-MLP are employed and evaluated for their performance. The results indicate high accuracy in predicting solar irradiance, with R-squared ($R^2$) scores close to unity for both train and test datasets. The impact of parameter integration on model performance is analyzed, revealing the significance of various parameters in enhancing predictive accuracy. Each model demonstrates strong performance across all parameters, consistently achieving MAE values below 6, RMSE values under 10, MBE within |2|, and nearly unity $R^2$ values. Upon removal of various solar parameters such as "Solar_Irradiance_Clear_Sky", "UVA", etc. from the datasets, the model's performance is significantly affected. This exclusion leads to considerable increases in MAE, reaching up to 82, RMSE up to 135, and MBE up to |7|. Among the models, KNN displays the weakest performance, with an $R^2$ of 0.7582546. Conversely, ANN exhibits the strongest performance, boasting an $R^2$ value of 0.9245877. Hence, the study concludes that Artificial Neural Network (ANN) performs exceptionally well, showcasing its versatility even under sparse data parameter conditions.

## 1. Introduction

In recent times, there has been a surge in the consumption of energy from conventional sources, leading to the rapid depletion of fossil fuel-based resources [1]. The increased utilization of conventional energy sources in recent decades has led to a rise in greenhouse gas emissions [2]. Additionally, escalating energy demands, driven by factors such as population growth, urbanization, and industrialization, have been observed in both developed and developing countries [3]. A share of this energy requirement can be satisfied through the utilization of renewable energy sources (RES), including hydropower, wind, and solar [4].

In the context of Nepal, hydropower is being used as a main source of energy for supplying electricity load demand and improving the system's reliability. However, the melting of snow in the Himalayas makes hydropower less feasible in the future thus increasing the need for alternative source of energy [5]. Among all the renewable energy sources (RESs), solar energy stands out as one of the most crucial and environmentally friendly resources. Photovoltaic (PV) energy plays a growing role in Nepal's energy mix due to its ease of installation and integration into building architecture, leveraging the country's favorable climatic conditions and excellent solar irradiance for grid-connected plants [6].

In contemporary times, numerous countries are exploring the integration of solar energy into their electricity grids. However, this poses additional challenges due to its intermittent and unpredictable nature giving rise to various issues, including voltage fluctuations and local power quality and stability concerns [7], [8] and challenges the continuity of the production/consumption balance [9]. Hence, a comprehensive understanding of the availability of solar radiation in a specific region is a crucial factor for the design and modeling of solar energy systems [10], [11], [12]. Consequently, predicting the output power of solar systems becomes essential for the effective functioning of the power grid and the optimal management of energy flows within the system [13]. In developing nations like Nepal, accurately measuring solar irradiance across various regions is challenging due to financial and technical constraints, underscoring the importance of creating diverse models and software for precise solar radiation estimation [14].

Solar radiation data is often unavailable for numerous potential sites [15], [16], primarily due to the high costs associated with the instruments [17], [18], [19], [20], issues related to sensor calibration, and equipment failures [21]. Hence, viable solutions to address this issue involve predicting solar radiation information based on alternative meteorological parameters [22].

## 2. Literature Review

In [23] various parameters such as sunshine hours, temperature, relative humidity, latitude, longitude, altitude, cloud cover, wind speed, transmittance, and pressure are used as input factors for solar power prediction. In the realm of solar radiation forecasting, [24] conducts a comprehensive analysis of machine learning techniques, comparing regression tree, random forest, and gradient boosting. Hybrid models, combining machine learning and physical models for enhanced performance, are highlighted. Cloud images from geostationary satellites, analyzed using the Heliosat method, serve short to long-term forecasting purposes in [25] and [26]. [27] compares a proposed method using Hidden Markov Models (HMM) and Generalized Fuzzy Models (GFM) with ANFIS and ANN, emphasizing its effectiveness for accurate solar radiation prediction, with sunshine duration as a pivotal parameter. [28] deploys ensemble machine learning (EML) algorithms to predict large-scale solar power generation in Eastern India based on meteorological parameters, achieving optimal results in RMSE and R2-score. [29] explores the influence of cloudiness on solar energy applications, extracting information from satellite-derived cloud images. [30] determines cloud motion from satellite images statistically, enabling solar radiation prediction over 30 minutes to 2 hours. [31] utilizes a deep neural network structure with satellite data for short-term solar irradiance forecasting, endorsing the global model as a cost-effective replacement for local models in 20 out of 25 locations. [32] introduces the UASIBS/KIER model, evaluating its performance against 35 ground observation stations over the Korean Peninsula, highlighting the reliability of daily aggregates. [33] discusses scaled persistence and ANN models for global solar radiation prediction, emphasizing their potential as alternatives to NWP models and underlining the importance of considering different seasons and weather conditions. [34] explores global solar radiation data sources for the Lerma Valley in Argentina, favoring LSASAF data for its better fit to ground measurements. [35] discusses the ESRA clear-sky model for automatic data processing, considering factors like the Linke turbidity factor and solar elevation. Shifting to weather forecasting. [36] presents a data-centric approach leveraging historical data, spatial interpolation, deep neural networks, and Gaussian Process regression to enhance predictions with variables including atmospheric pressure, temperature, dew point, and winds. In [37], a hybrid neural model (MLP and RBF) for weather forecasting in Saudi Arabia, incorporating features like average dew point, temperature, relative humidity, precipitation, wind speed, and cloudiness, is found superior to individual feedforward neural networks. [38] notes a study extracting observed and 6-day forecasted weather data from NOAA, revealing biases and overestimation except for wind speed, with relative humidity playing a pivotal role in energy forecasting. [39] introduces a rain prediction system using real-time data, achieving 87.90% accuracy with machine learning algorithms, specifically random forest classification. [40] identifies variables for precise solar radiation prediction using WEKA software, applied across 26 Indian locations to enhance ANN model accuracy. [41] models global solar radiation in Algeria using temperature and relative humidity inputs, employing the LM feed-forward backpropagation algorithm. [42] constructs multi-layered neural networks for solar energy potential forecasting in 195 Nigerian cities, relying on NASA data over a 10-year period and evaluating reliability based on the correlation coefficient.

[43] developed ANN models (ANN-I5, ANN-I4, and ANN-I3) for assessing solar potential in Himachal Pradesh, favoring the ANN-I5 model with all five input parameters for accurate solar radiation prediction. [44] introduced a novel approach using AI techniques (ANN and ANFIS) for effective forecasting of mean hourly global solar radiation. [45] proposed an ANN model using ground and satellite data for solar radiation prediction in Gran Canaria, emphasizing the superiority of annual pixel selection. [46] discussed the use of models like the Angström–Prescott equation for estimating global solar radiation, highlighting challenges and assessment parameters. [47] explored ANN models for weather forecasting, focusing on predicting maximum temperature for 365 days and claiming applicability to other weather factors. [48] used ANN and physical models to forecast solar radiation across 12 Turkish cities, obtaining RMSE values of 91 W/m2 for training cities and 125 W/m2 for testing cities. [49] utilized ANN to estimate hourly and daily values of the diffuse fraction (KD) in Egypt, incorporating various meteorological inputs.

[50] proposes a new intra-hour solar irradiance forecast method using optimized pattern recognition for improved predictions of GHI and DNI, particularly in capturing large ramps. [51] discusses a semi-empirical approach, utilizing gradient boosting and k-nearest neighbors regression for probabilistic solar power forecasting, beneficial for the GEFCom2014 competition and operational forecasts for solar power plants within short to medium timescales. [52] explores various methods, including neural networks and genetic algorithms, for modeling and forecasting non-linear processes in renewable energy grids, evaluating their performance. [53] assesses SVM and KNN performance, suggesting SVM's effectiveness with small samples and highlighting KNN's sensitivity to the length of the training dataset, potentially achieving higher accuracy with sufficient samples.

[54] introduces a hybrid CNN-LSTM model, tested with PV power data in Busan, Korea, aiming to improve short-term PV forecasting accuracy. [55] utilizes deep learning algorithms (Deep Belief Networks, Auto Encoder, LSTM) in 21 solar power cases, demonstrating superior forecasting performance compared to traditional methods. [56] presents a "modified multi-step CNN-stacked LSTM" machine learning model for highly accurate solar irradiance prediction (GHI and POA), combining CNNs for feature extraction with stacked LSTMs for sequence prediction. [57] addresses challenges in integrating distributed energy resources, proposing an advanced one-day ahead solar power forecasting model using a simplified LSTM algorithm, achieving an average RMSE of 0.512.

[58] introduces a random forest algorithm for solar power plant energy output forecasting, achieving 8% mean prediction error on a Russian plant. [59] evaluates smart persistence, ANN,

and random forest models for solar components, favoring random forest for h+1 to h+6 time horizons. [51] successfully applies gradient boosting and k-nearest neighbors for accurate probabilistic solar power forecasting in GEFCom2014. [60] proposes a validated random forest for superior day-ahead PV power output predictions, emphasizing short-term accuracy for grid integration efficiency. [61] presents a robust random forest algorithm, combining PCA, K-means, and Differential Evolution Grey Wolf Optimizer for accurate and stable photovoltaic power generation prediction, suggesting enhancements for efficient variable screening and parameter selection.

[62] proposes an XGBoost-based probabilistic model for solar irradiance forecasting with superior accuracy and reduced training time. [63] introduces an effective Light GBM-XGBoost model for predicting PV power output, demonstrating superior accuracy and adaptability. [64] presents an XGBoost-based load forecasting algorithm for behind-the-meter solar PV in South Korea, improving accuracy and economic efficiency. [65] utilizes XGBoost regression for solar power prediction, outperforming SVM with an RMSE of 6.63% in Johannesburg, aiming to enhance grid stability.

## 3. Methodology

*3.1. Study Area*

This study centers on solar radiation prediction within five distinct regions in Nepal. These regions were thoughtfully selected for their distinctive geographical characteristics and their significance in solar energy applications. The chosen areas comprise:

- **Kathmandu:** Kathmandu, the capital city of Nepal, is located at a latitude of approximately 27.7° and a longitude of approximately 86.366°, with an altitude of 1,337 meters above sea level. Situated in the Warm Temperate Zone, Kathmandu experiences average summer temperatures ranging from 28° to 30°C and winter temperatures between 3 and 10.1°C. Given its geographical location and favorable weather conditions, Kathmandu is well-suited for harnessing solar power.

- **Nuwakot:** Nuwakot, located at a latitude of 27.9265° and a longitude of 85.2477°, stands at an altitude of 1,022 meters above sea level with an average summer temperature ranging from 28° to 34°C and winter temperatures between 5 and 23°C. Nuwakot's climate is intricately shaped by its altitude and geographical features, leading to fluctuations in temperature and precipitation. Remarkably, it houses Nepal's largest solar power plant, strategically installed across six locations within the Devighat Hydropower Station premises, which is owned by the NEA (Nepal Electricity Authority).

- **Mustang:** Mustang, commonly known as the "Forbidden Kingdom," is situated at a latitude of 28.9814° and a longitude of 83.8606°, boasting an altitude of 4,023 meters above sea level with an average annual temperature of 10.9 °C. The weather conditions in Mustang are distinctly shaped by its elevated altitude, resulting in exceptional temperature ranges. Given its high altitude and predominantly clear weather conditions, it implies that Mustang could be well-suited for harnessing solar power.

- **Dhangadhi:** Located in the far-western region, Dhangadhi has a latitude of 28.7016° and a longitude of 80.5898°, accompanied by an altitude of 109 meters above sea level with an average annual temperature of 22.84°C. The city's relatively low altitude contributes to temperature variations throughout different seasons. Dhangadhi's geographical position and the likelihood of ample sunlight exposure imply its potential suitability for solar power.

- **Taplejung:** Situated in eastern Nepal, Taplejung exhibits varied weather patterns shaped by its specific latitude of 27.5875° and longitude of 87.8218°, accompanied by an elevation of 1,442 meters above sea level with an average annual temperature of 27.91°C. The elevated altitude of Taplejung, coupled with its promising exposure to sunlight, suggests its potential suitability for the harnessing of solar power.

*3.2. Dataset Description*

The research utilized a dataset covering an extensive timeframe of almost ten years, starting from January 1st, 2014, to September 31st, 2023. This dataset includes a diverse set of meteorological variables and irradiance data obtained from two reputable sources: CERES SYN1deg (NASA's CERES Satellite) and MERRA-2 (NASA's Modern-Era Retrospective Analysis for Research and Applications, Version 2).

The parameters of the dataset are:
- **Solar_Irradiance:** The total shortwave (solar) radiation reaching the Earth's surface under all sky conditions. This includes both direct and diffuse components of sunlight ($Wh/m^2$).
- **Solar_Irradiance_Clear_Sky:** The amount of solar radiation received per unit area on the Earth's surface under clear sky conditions ($Wh/m^2$).
- **Photo_Radiation:** The total Photosynthetically Active Radiation (PAR) reaching the Earth's surface under all sky conditions ($W/m^2$).
- **Photo_Radiation_Clear_Sky:** The total Photosynthetically Active Radiation (PAR) reaching the Earth's surface under clear sky conditions ($W/m^2$).
- **UVA:** Ultraviolet A (UVA) irradiance reaching the Earth's surface under all sky conditions ($W/m^2$).
- **UVB:** Ultraviolet B (UVB) irradiance reaching the Earth's surface under all sky conditions ($W/m^2$).
- **UV_Index:** A dimensionless index representing the strength of ultraviolet (UV) radiation reaching the Earth's surface under all sky conditions.
- **T2M:** The air temperature at a height of 2 meters above the Earth's surface (C).
- **DEW2M:** The temperature at which air becomes saturated with moisture and dew or frost begins to form at a height of 2 meters (C).
- **T2MWET:** The temperature at which air cools to saturation (100% relative humidity) with the help of water evaporation at a height of 2 meters(C).
- **SH2M:** Specific Humidity at 2 Meters (g/kg)
- **RH2M:** Relative Humidity at 2 Meters (%)
- **Precipitation:** The corrected precipitation rate, representing the amount of rainfall per unit area per hour (mm/hour).
- **SP:** Surface Pressure (kPa)
- **WS10M:** Wind Speed at 10 Meters (m/s)
- **WD10M:** Wind Direction at 10 Meters (Degrees)

*3.3. Data Split for Model Testing*

The dataset is split into two parts:
- **Data1: January 2014 to December 2022**

    This dataset is further split into train and test set with test size of 0.2(20% of total data) and 80% of the data is used to train the machine learning model.
- **Data2: January 2023 to September 2023**

    The machine learning model trained with the above dataset is now used to predict this dataset and the performance of the model is evaluated.

*3.4. Random Forest*

The Random Forest (RF) stands as an ensemble model comprised of multiple decision tree models, as illustrated in Figure 1. In the RF algorithm, each tree is trained on a randomly selected subset of the dataset, generating individual predictions. Particularly adept at handling high-dimensional datasets, RF overcomes limitations associated with traditional regression analysis methods. It employs the bagging method at its core, sampling multiple datasets, training multiple decision trees, and aggregating their results through a majority vote in classification problems or computing the mean value in regression problems. The Random Forest's performance hinges on its hyper parameters, necessitating hyper parameter tuning through k-fold validation to select values yielding minimal error values [66], [67].

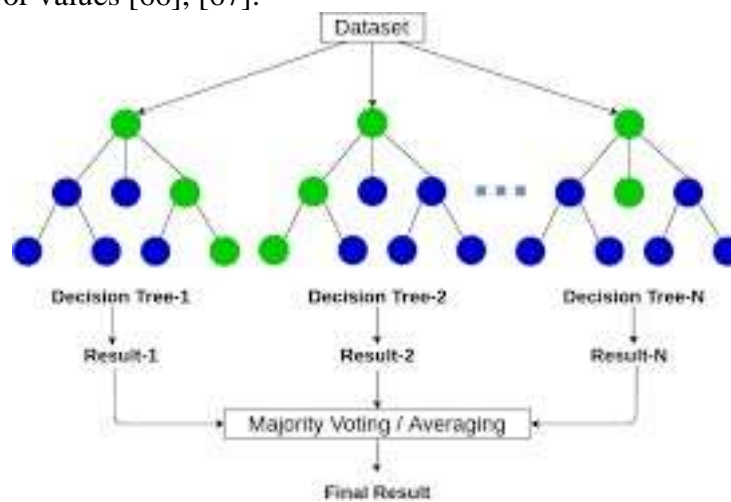

Figure 1: Illustration of random forest tree

*3.5. Extreme Gradient Boosting (XGBoost)*

XGBoost, or Extreme Gradient Boosting, introduced in 2011 by Tianqi Chen and Carlos Guestrin and continually refined through subsequent research [68], is a learning framework centered around Boosting Tree models. Diverging from traditional Boosting Tree models, XGBoost utilizes a second-order Taylor expansion on the loss function, incorporating CPU multithreading for parallel computing. It addresses challenges in distributed training by avoiding reliance solely on first-derivative information. Noteworthy for its performance and versatility in handling regression and classification tasks, XGBoost employs various strategies to combat overfitting.

## 3.6. K-Nearest Neighbors (KNN)

KNN, a well-established statistical method with a 40-year history in pattern recognition, has found application in text categorization, including in the Reuters benchmark dataset. The KNN approach involves calculating distances between new samples and training data, identifying the K nearest neighbors, and determining the category of the new sample based on the categories of these neighbors. If all neighbors belong to the same category, the new sample is assigned to that category; otherwise, certain rules are applied to score each potential category and determine the new sample's category accordingly [69]. The KNN (K Nearest Neighbor) algorithm is a nonparametric technique employed for both classification and regression tasks [70]. The K Nearest Neighbor (KNN) method has found extensive applications in the realms of data mining and machine learning, primarily owing to its straightforward implementation and notable performance [71].

## 3.7. Long Short-Term Memory (LSTM) neural network

LSTMs (Long Short-Term Memory networks) represent an advanced neural network modeling approach, offering an enhanced version of RNNs that excels at capturing long-term dependencies within sequential time series data. LSTMs effectively counter the challenges of vanishing and exploding gradients often encountered in conventional RNN architectures, rendering them an excellent choice for modeling and forecasting sequential time series data. This model employs a gate control mechanism to manage information flow, allowing it to selectively decide how much incoming data to retain for each time step. The fundamental LSTM unit comprises three key control gates: an input gate, an output gate, and a forgetting gate [72].

## 4. Results & Discussions

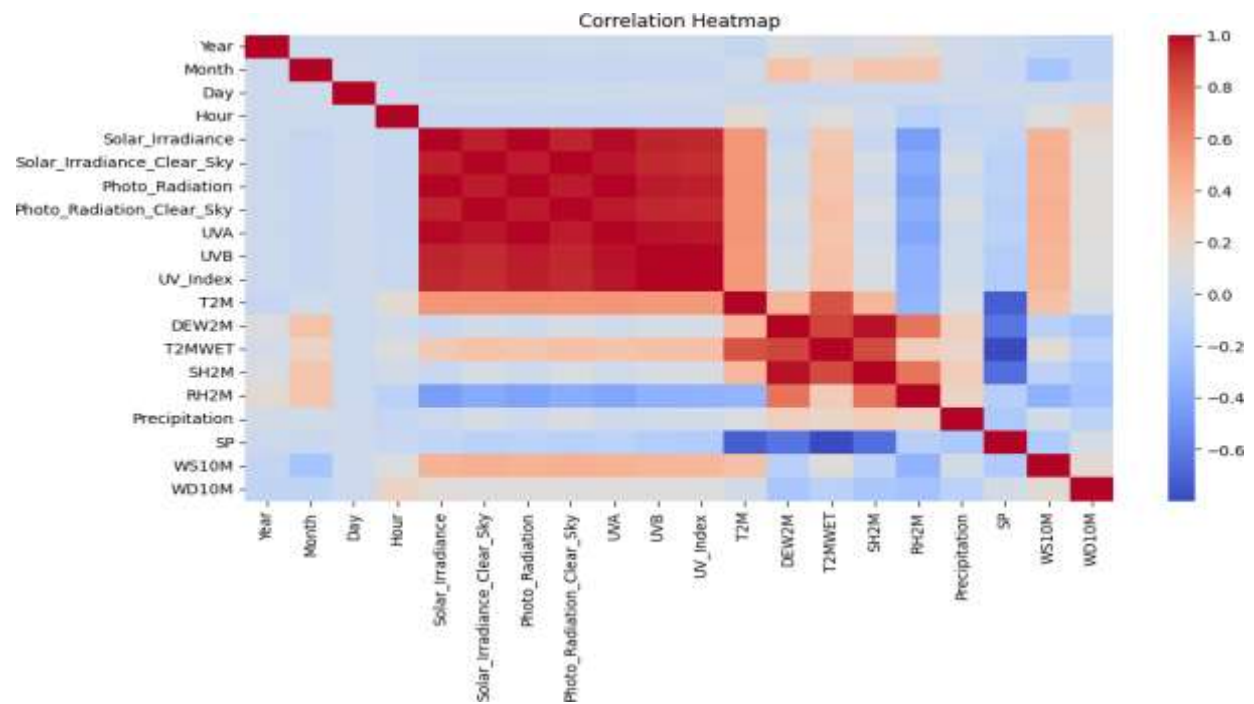

Figure 2: Heatmap showing correlation of solar irradiance with various parameters

Figure 2 depicts a heatmap illustrating the correlation of solar irradiance with diverse parameters. The analysis concludes that "Solar_Irradiance" shows a distinct correlation with solar-related factors such as "Solar_Irradiance_Clear_Sky," "Photo_Radiation," "UVA," and others. Moreover, a significant association exists between our focal parameter and meteorological variables like "WS10M" and "T2M."

### 4.1. Model Performance on Training Dataset Data1

The dataset data1 is split into training and testing sets and used to train each of the machine learning models described in sections 3.4-3.7 A scatter plot is visualized for the predicted and actual Solar Irradiance Value for every machine learning models(KNN, XGBoost, Random Forest), Deep Learning model (LSTM), and Artificial Neural Network(MLP).

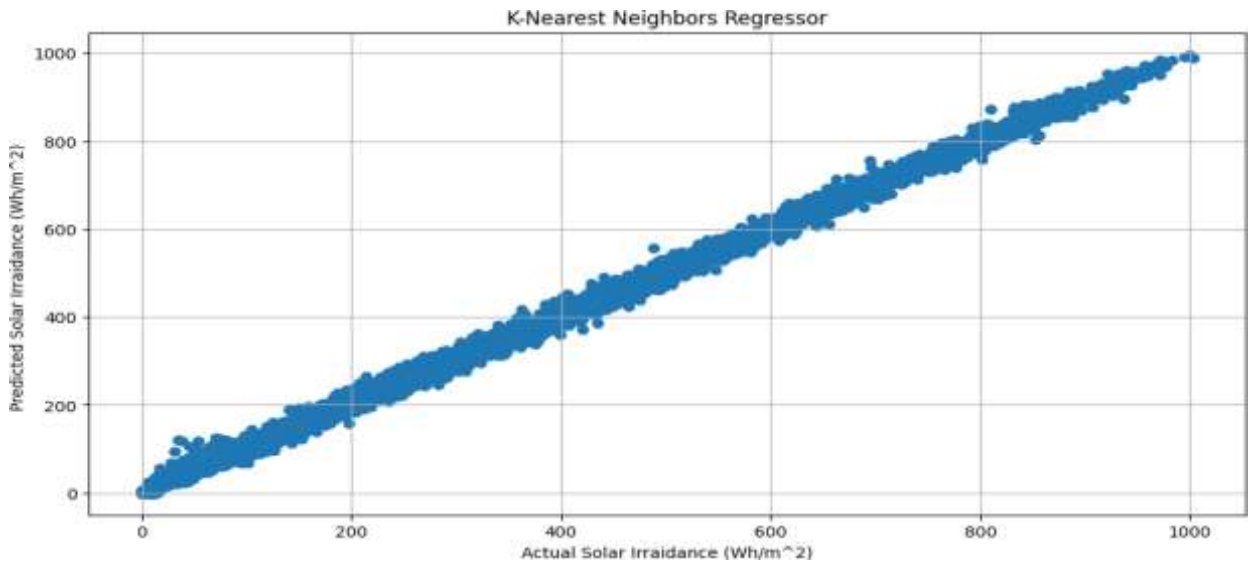

Figure 3: Scatter plot of predicted and actual solar irradiance using KNN algorithm

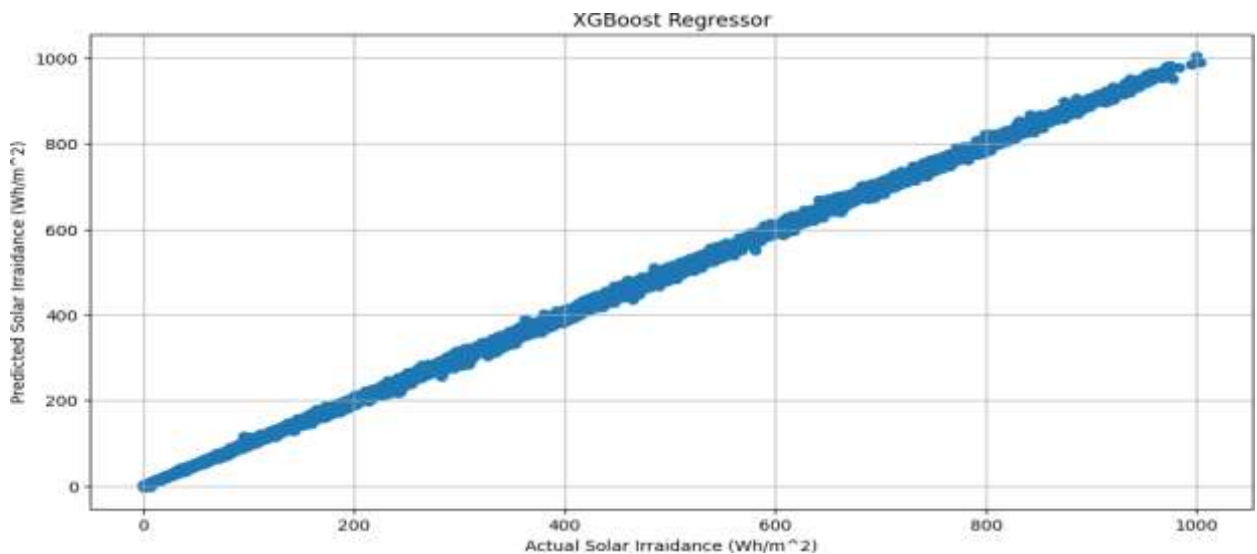

Figure 4: Scatter plot of predicted and actual solar irradiance using XGBoost Algorithm

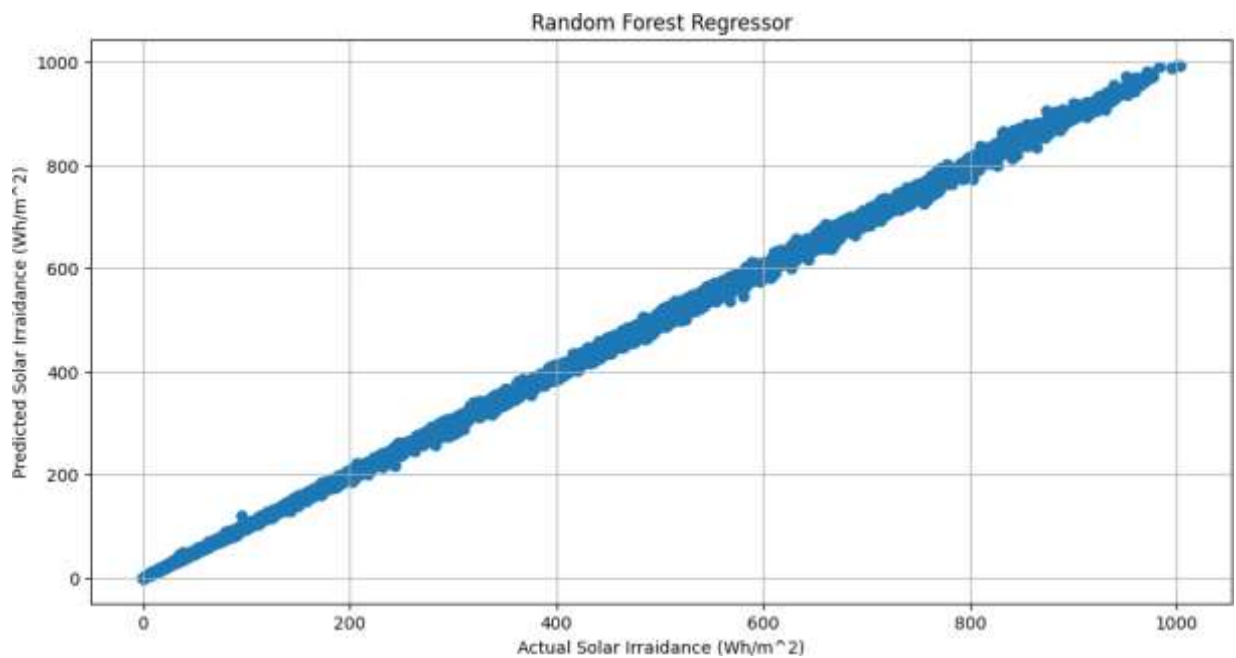

Figure 5: Scatter plot of predicted and actual solar irradiance using random forest algorithm

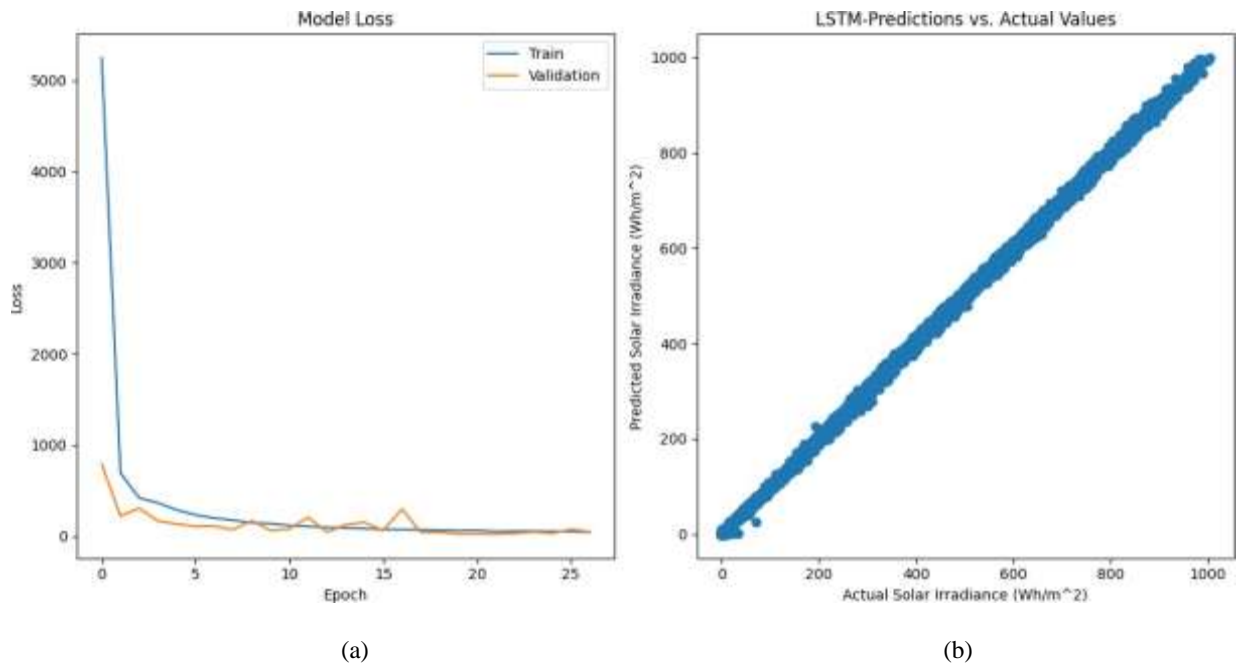

(a)            (b)

Figure 6: (a) performance of LSTM – deep learning model showcasing model loss vs epoch

(b) Scatter plot of predicted and actual solar irradiance using LSTM algorithm

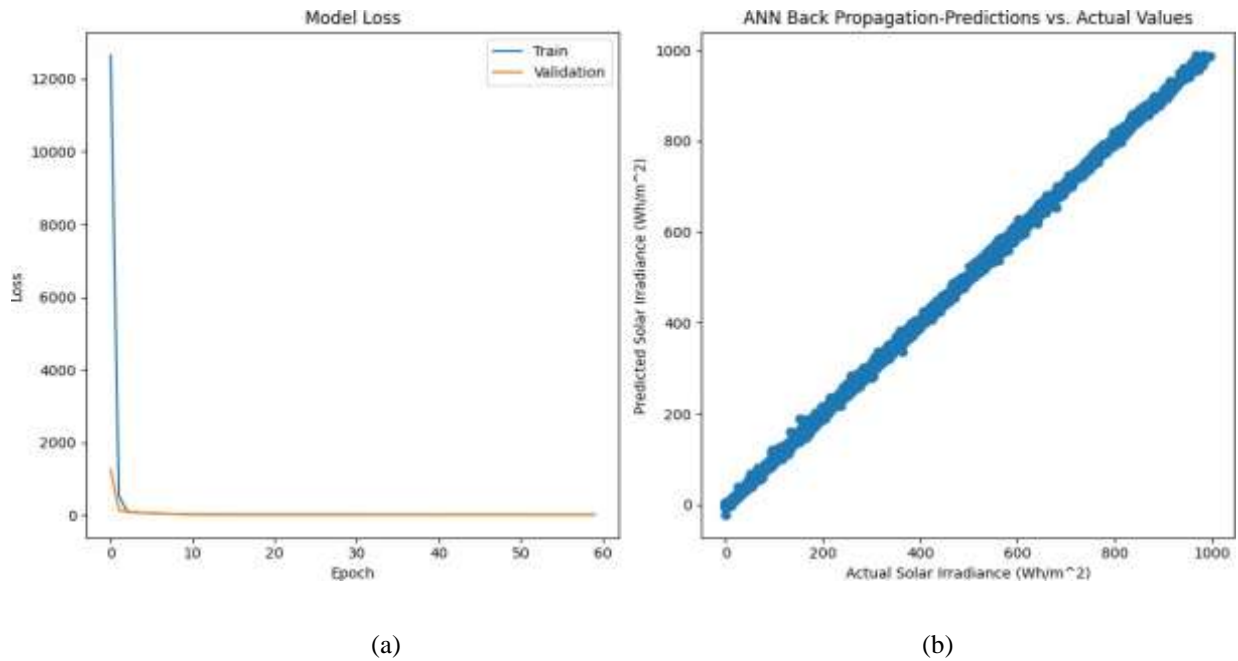

(a)                  (b)

Figure 7: (a) performance of MLP - ANN showcasing model loss vs epoch

(b) Scatter plot of predicted and actual solar irradiance using MLP

Scatter plot from figure 3-7, shows the performance analysis of actual data and predicted data for each model used and indicates that a regression line can easily fit the scatter plot with a positive slope straight line showing that our model performs well. For deep learning and ANN model, model loss vs epoch is also plotted depicting that the model loss is decreasing per epoch.

Then, the model performance is evaluated for the test set using an R-squared ($R^2$) score for all the geographical study locations as shown in table 1. It is found that every model performs well with $R^2$ score value of almost unity.

Table 1: $R^2$ Score of different models for dataset1

| Model Name Location | LSTM | Random Forest | KNN | XGB | ANN MLP |
|---|---|---|---|---|---|
| Kathmandu | 0.9994070 | 0.9996082 | 0.9989602 | 0.9996691 | 0.9998680 |
| Nuwakot | 0.9993468 | 0.9996090 | 0.9990477 | 0.9996670 | 0.9996706 |
| Dhangadhi | 0.9997086 | 0.9997487 | 0.9989415 | 0.9997634 | 0.9998117 |
| Mustang | 0.9981421 | 0.9996363 | 0.9987487 | 0.9996889 | 0.9997685 |
| Taplejung | 0.9981720 | 0.9995538 | 0.9988035 | 0.9995736 | 0.9997182 |

### 4.2 Model Performance on Data2

The model that we have trained using the data 2014-2022 is now used to predict unknown data of Jan-Sept 2023. The predicted data and actual data is plotted for each model.

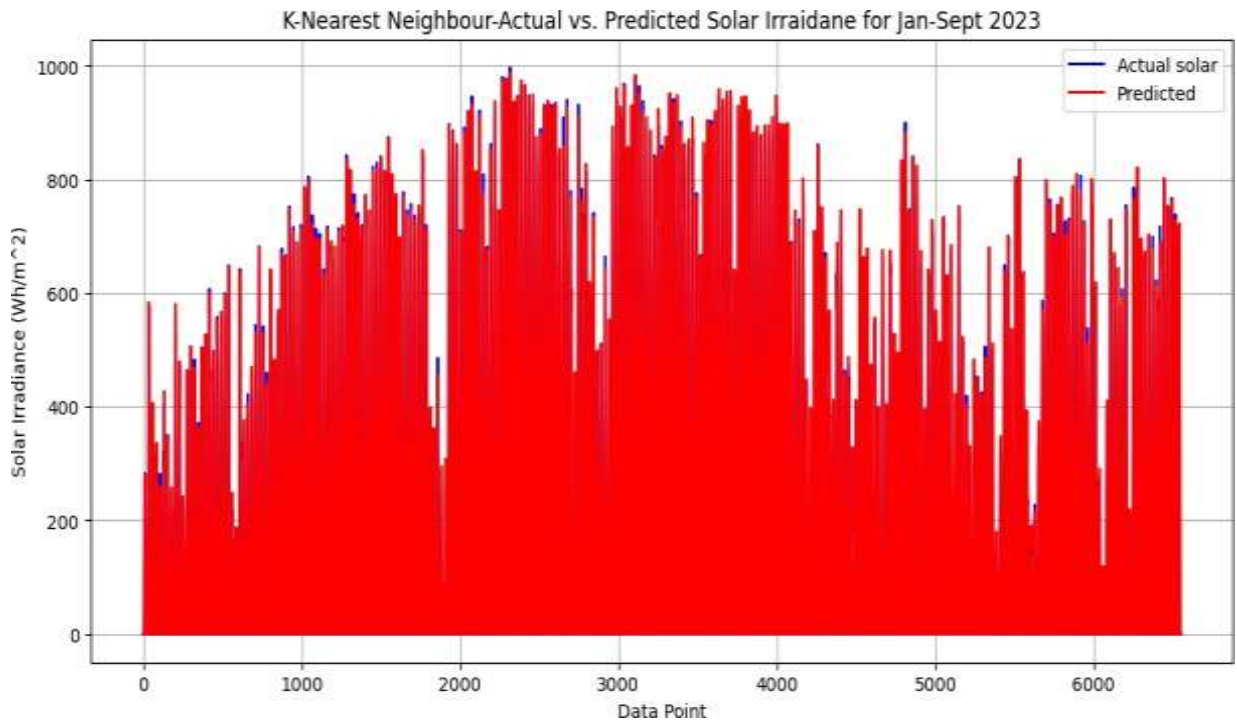

Figure 8: Predicted vs actual solar irradiance for KNN algorithm

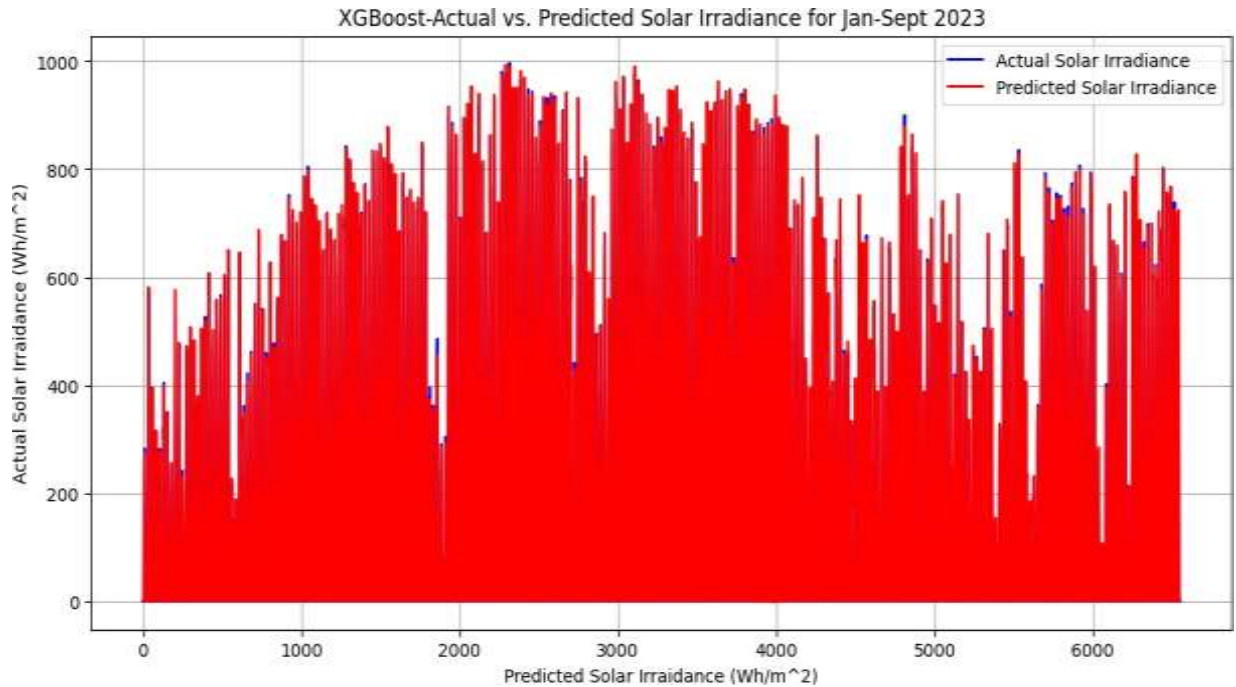

Figure 9: Predicted vs actual solar irradiance for XGBoost algorithm

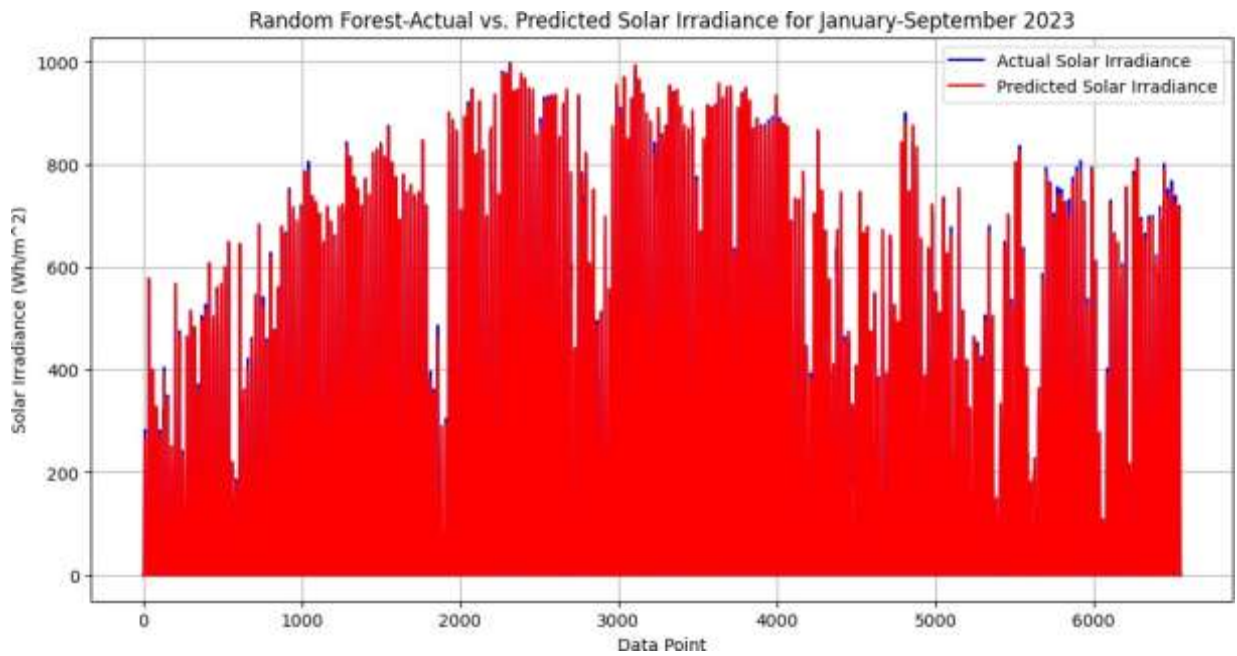

Figure 10: Predicted vs actual solar irradiance for random forest algorithm

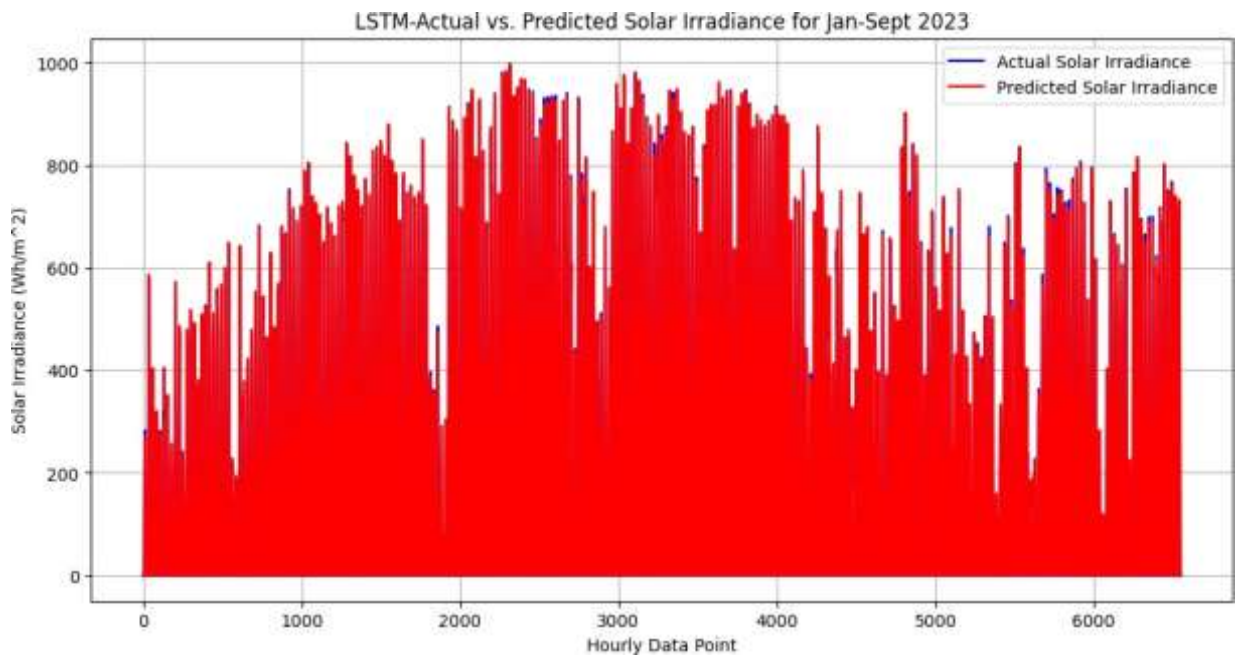

Figure 11: Predicted vs actual solar irradiance for LSTM - deep learning model

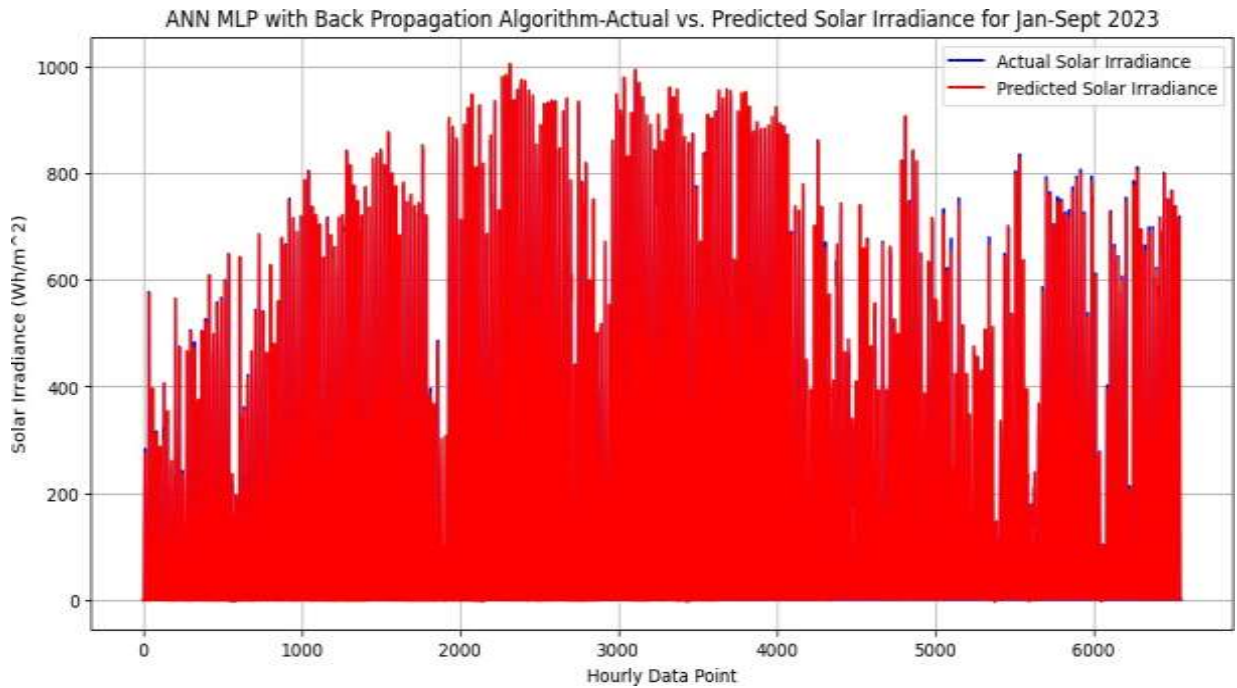

Figure 12: Predicted vs actual solar irradiance for ANN-MLP model with higher accuracy.

Figure 8-12 shows the performance analysis of predicted and actual solar irradiance for various models. The red line indicates the predicted solar irradiance and blue line indicating the actual solar irradiance and we can see that almost every predicted point fits the actual point. The model performance is also evaluated using an R-squared ($R^2$) score for all the geographical study locations as shown in table 2. It is found that every model performs well with $R^2$ score value of almost unity even with the unknown Dataset2, indicating that the model can be used to forecast solar irradiance with higher accuracy.

Then, the model performance is evaluated for the test set using an R-squared ($R^2$) score for all the geographical study locations as shown in table 1. It is found that every model performs well with $R^2$ score value of almost unity.

Table 2: $R^2$ Score of different models for dataset2

| Model Name Location | LSTM | Random Forest | KNN | XGB | ANN MLP |
|---|---|---|---|---|---|
| Kathmandu | 0.9991981 | 0.9995326 | 0.9989865 | 0.9995806 | 0.9998289 |
| Nuwakot | 0.9989819 | 0.9995437 | 0.9990563 | 0.9995468 | 0.9997211 |
| Dhangadhi | 0.9996120 | 0.9997139 | 0.9989449 | 0.9996966 | 0.9997958 |
| Mustang | 0.9980786 | 0.9996130 | 0.9988141 | 0.9996078 | 0.9997306 |
| Taplejung | 0.9969656 | 0.9995016 | 0.9987566 | 0.9993963 | 0.9995574 |

## 4.3 Impact of parameter integration on model performance

A comparative analysis of various models is performed using metrics like MAE, RMSE, MBE, and $R^2$ taking Dhangadhi as a typical example.

Table 3: Metrics for model evaluation of Dhangadhi dataset2

| Model Name Parameters | LSTM | Random Forest | KNN | XGB | ANN MLP |
|---|---|---|---|---|---|
| MAE | 5.8803 | 2.2622 | 4.8004 | 2.5526 | 2.3955 |
| RMSE | 9.8950 | 4.6633 | 8.9559 | 4.8026 | 3.9259 |
| MBE | -1.9796 | 0.1771 | -0.8351 | -0.4297 | 0.0605 |
| $R^2$ | 0.9987120 | 0.9997139 | 0.9989449 | 0.9996966 | 0.9997973 |

Table 3 shows the metrics for model evaluation of Dhangadhi dataset2. Here, the dataset2 is trained with different models using all the parameters available in the data and we can see that every model performs well with MAE< |2| and with $R^2$ value almost unity.

Table 4: Metrics for model evaluation of Dhangadhi dataset2 without Solar_Irradiance_Clear_Sky

| Model Name Parameters | LSTM | Random Forest | KNN | XGB | ANN MLP |
|---|---|---|---|---|---|
| MAE | 3.6660 | 2.5365 | 4.9681 | 2.8375 | 3.0179 |
| RMSE | 6.4275 | 5.0585 | 8.7778 | 5.2830 | 5.2646 |
| MBE | 0.0328 | 0.0542 | -0.3814 | -0.5174 | -0.3928 |
| $R^2$ | 0.9994565 | 0.9996634 | 0.9989864 | 0.9996328 | 0.9996354 |

Table 4 shows metrics for model evaluation of Dhangadhi dataset2 without Solar_Irradiance_Clear_Sky. Here, the dataset2 is trained with different models using all the parameters except "Solar_Irradiance_Clear_Sky" and we can see that every model performs well with MAE< |0.6| and with $R^2$ value almost unity.

Table 5: Metrics for model evaluation of Dhangadhi dataset2 without any solar parameters

| Model Name Parameters | LSTM | Random Forest | KNN | XGB | ANN MLP |
|---|---|---|---|---|---|
| MAE | 40.9883 | 36.5331 | 82.9040 | 43.5091 | 39.2850 |
| RMSE | 78.7979 | 76.0982 | 135.5671 | 80.3954 | 75.7175 |
| MBE | 1.1445 | 1.1253 | -7.0125 | 2.9191 | -4.5789 |
| $R^2$ | 0.9183267 | 0.9238273 | 0.7582546 | 0.9149818 | 0.9245877 |

Table 5 shows metrics for model evaluation of Dhangadhi dataset2 without any solar parameters. All the solar parameters like "Solar_Irradiance_Clear_Sky", "UVA", etc. are removed from the data while training and the model is evaluated with different metrics as before. We can see that removal of parameters directly affect the model performance such that MAE increases to up to 82, RMSE to

135, MBE to |7|. KNN performs worst with $R^2$ value of 0.7582546. ANN performs the best with 0.9245877 $R^2$ value.

## 5. Conclusion

In the context of Nepal, we have evaluated machine learning models for predicting solar irradiance at different locations. Our results emphasize that, with sufficient data and parameters, every model can effectively forecast solar irradiance, achieving an impressive $R^2$ value close to unity. The reduction of data parameters adversely affects model performance, leading to increased errors and a decrease in the $R^2$ score. Notably, among all models, Artificial Neural Network (ANN) demonstrates superior performance, even under sparse data parameter conditions, with an $R^2$ value of 0.9245877, showcasing its versatility across various use cases.

**References**

[1]  Y. Khan and F. Liu, "Consumption of energy from conventional sources a challenge to the green environment: evaluating the role of energy imports, and energy intensity in Australia," *Environmental Science and Pollution Research*, vol. 30, no. 9, pp. 22712–22727, Feb. 2023, doi: 10.1007/s11356-022-23750-x.

[2]  L. Jeffry, M. Y. Ong, S. Nomanbhay, M. Mofijur, M. Mubashir, and P. L. Show, "Greenhouse gases utilization: A review," *Fuel*, vol. 301, Oct. 2021, doi: 10.1016/j.fuel.2021.121017.

[3]  M. Shahbaz, R. Sbia, and T. Afza, "The Effect of Urbanization, Affluence and Trade Openness on Energy Consumption: A Time Series Analysis in Malaysia," 2015.

[4]  L. S. Paraschiv and S. Paraschiv, "Contribution of renewable energy (hydro, wind, solar and biomass) to decarbonization and transformation of the electricity generation sector for sustainable development," *Energy Reports*, vol. 9, pp. 535–544, Sep. 2023, doi: 10.1016/j.egyr.2023.07.024.

[5]  S. R. Bajracharya, P. K. Mool, and B. R. Shrestha, "The impact of global warming on the glaciers of the Himalaya."

[6]  N. Kharipati, N. Bhattrai, A. Kumar Jha, B. R. Tiwari, and N. Bhattarai, "Performance analysis of a 100 kWp grid connected Solar Photovoltaic Power Plant in Performance analysis of a 100 kWp grid connected Solar Photovoltaic Power Plant in Kharipati, Bhaktapur, Nepal," 2017, [Online]. Available: https://www.researchgate.net/publication/341592112

[7]  IEEE Electron Devices Society, Annual IEEE Computer Conference, Calif. IEEE Photovoltaic Specialists Conference 33 2008.05.11-16 San Diego, and Calif. PVSC 33 2008.05.11-16 San Diego, *33rd IEEE Photovoltaic Specialists Conference, 2008 PVSC '08 ; 11-16 May 2008, San Diego, California ; conference proceedings*.

[8]  D. Anderson and M. Leach, "Harvesting and redistributing renewable energy: On the role of gas and electricity grids to overcome intermittency through the generation and storage of hydrogen," *Energy Policy*, vol. 32, no. 14, pp. 1603–1614, Sep. 2004, doi: 10.1016/S0301-4215(03)00131-9.

[9]  V. Lara-Fanego, J. A. Ruiz-Arias, D. Pozo-Vázquez, F. J. Santos-Alamillos, and J. Tovar-Pescador, "Evaluation of the WRF model solar irradiance forecasts in Andalusia (southern Spain)," *Solar Energy*, vol. 86, no. 8. pp. 2200–2217, Aug. 2012.

[10] X. Vallvé and Ostbayerisches Technologie-Transfer-Institut Bereich Erneuerbare Energien, *5th European Conference PV Hybrid and Mini Grid Tarragona, Spain, April 29th/30th, 2010*.

[11] S. Kaplanis, J. Kumar, and E. Kaplani, "On a universal model for the prediction of the daily global solar radiation," *Renew Energy*, vol. 91, pp. 178–188, Jun. 2016, doi: 10.1016/j.renene.2016.01.037.

[12] A. D. Naser, "An estimation of the monthly average daily diffuse solar radiation on horizontal surfaces over libya," *Energy Sources, Part A: Recovery, Utilization and Environmental Effects*, vol. 33, no. 4, pp. 317–326, Jan. 2011, doi: 10.1080/15567030902967884.

[13] M. Paulescu, E. Paulescu, P. Gravila, and V. Badescu, "Weather Modeling and Forecasting of PV Systems Operation," *Green Energy and Technology*, vol. 103, 2013, doi: 10.1007/978-1-4471-4649-0.

[14] F. Limbu, B. R. Tiwari, U. Joshi, J. Regmi, I. B. Karki, and K. N. Poudyal, "Comparative study of solar flux using different empirical models at low land urban industrial zone of Biratnagar Nepal," *Himalayan Physics*, pp. 100–109, Jun. 2023, doi: 10.3126/hp.v10i1.55286.

[15] V. H. Quej, J. Almorox, M. Ibrakhimov, and L. Saito, "Estimating daily global solar radiation by day of the year in six cities located in the Yucatán Peninsula, Mexico," *J Clean Prod*, vol. 141, pp. 75–82, Jan. 2017, doi: 10.1016/j.jclepro.2016.09.062.

[16] T. Khatib, A. Mohamed, K. Sopian, and M. Mahmoud, "Solar energy prediction for Malaysia using artificial neural networks," *International Journal of Photoenergy*, vol. 2012, 2012, doi: 10.1155/2012/419504.

[17] Bangladesh. F. of E. and E. E. Khulna University of Engineering & Technology, Institute of Electrical and Electronics Engineers. Bangladesh Section. EDS/SSCS Chapter, and Institute of Electrical and Electronics Engineers, *2013 International Conference on Electrical Information and Communication Technology : EICT 2013 : 13-15 February 2014 : Khulna, Bangladesh*.

[18] Z. Ramedani, M. Omid, and A. Keyhani, "Modeling solar energy potential in a tehran province using artificial neural networks," *Int J Green Energy*, vol. 10, no. 4, pp. 427–441, Apr. 2013, doi: 10.1080/15435075.2011.647172.

[19] Md. A. R. Ahad, M. Turk, Institute of Electrical and Electronics Engineers, IEEE Computer Society, and Center for Natural Science & Engineering Research, *5th International Conference on Informatics, Electronics and Vision : 13-14 May, 2016, Dhaka, Bangladesh*.

[20] M. Şahin, "Comparison of modelling ANN and ELM to estimate solar radiation over Turkey using NOAA satellite data," *Int J Remote Sens*, vol. 34, no. 21, pp. 7508–7533, 2013, doi: 10.1080/01431161.2013.822597.

[21] J. Díaz-Gómez, A. Parrales, A. Álvarez, S. Silva-Martínez, D. Colorado, and J. A. Hernández, "Prediction of global solar radiation by artificial neural network based on a meteorological environmental data," *Desalination Water Treat*, vol. 55, no. 12, pp. 3210–3217, Sep. 2015, doi: 10.1080/19443994.2014.939861.

[22] D. V. Siva Krishna Rao K, M. Premalatha, and C. Naveen, "Models for forecasting monthly mean daily global solar radiation from in-situ measurements: Application in Tropical Climate, India," *Urban Clim*, vol. 24, pp. 921–939, Jun. 2018, doi: 10.1016/j.uclim.2017.11.004.

[23] R. C. Deo, X. Wen, and F. Qi, "A wavelet-coupled support vector machine model for forecasting global incident solar radiation using limited meteorological dataset," *Appl Energy*, vol. 168, pp. 568–593, Apr. 2016, doi: 10.1016/j.apenergy.2016.01.130.

[24] C. Voyant *et al.*, "Machine learning methods for solar radiation forecasting: A review,"

*Renewable Energy*, vol. 105. Elsevier Ltd, pp. 569–582, 2017. doi: 10.1016/j.renene.2016.12.095.

[25] E. Lorenz, A. Hammer, and D. Heinemann, "Short term forecasting of solar radiation based on satellite data."

[26] D. Heinemann, E. Lorenz, and M. Girodo, "Forecasting of Solar Radiation."

[27] S. Bhardwaj *et al.*, "Estimation of solar radiation using a combination of Hidden Markov Model and generalized Fuzzy model," *Solar Energy*, vol. 93, pp. 43–54, Jul. 2013, doi: 10.1016/j.solener.2013.03.020.

[28] D. Chakraborty, J. Mondal, H. B. Barua, and A. Bhattacharjee, "Computational solar energy – Ensemble learning methods for prediction of solar power generation based on meteorological parameters in Eastern India," *Renewable Energy Focus*, vol. 44, pp. 277–294, Mar. 2023, doi: 10.1016/j.ref.2023.01.006.

[29] "SHORT-TERM FORECASTING OF SOLAR RADIATION: A STATISTICAL APPROACH USING SATELLITE DATA¨," 1999. [Online]. Available: www.elsevier.com/locate/solener

[30] R. Marquez, H. T. C. Pedro, and C. F. M. Coimbra, "Hybrid solar forecasting method uses satellite imaging and ground telemetry as inputs to ANNs," *Solar Energy*, vol. 92, pp. 176–188, Jun. 2013, doi: 10.1016/j.solener.2013.02.023.

[31] J. Lago, K. De Brabandere, F. De Ridder, and B. De Schutter, "Short-term forecasting of solar irradiance without local telemetry: A generalized model using satellite data," *Solar Energy*, vol. 173, pp. 566–577, Oct. 2018, doi: 10.1016/j.solener.2018.07.050.

[32] C. K. Kim, H. G. Kim, Y. H. Kang, and C. Y. Yun, "Toward Improved Solar Irradiance Forecasts: Comparison of the Global Horizontal Irradiances Derived from the COMS Satellite Imagery Over the Korean Peninsula," *Pure Appl Geophys*, vol. 174, no. 7, pp. 2773–2792, Jul. 2017, doi: 10.1007/s00024-017-1578-y.

[33] C. Voyant, P. Haurant, M. Muselli, C. Paoli, and M. L. Nivet, "Time series modeling and large scale global solar radiation forecasting from geostationary satellites data," *Solar Energy*, vol. 102, pp. 131–142, Apr. 2014, doi: 10.1016/j.solener.2014.01.017.

[34] L. Ramirez Camargo and W. Dorner, "Comparison of satellite imagery based data, reanalysis data and statistical methods for mapping global solar radiation in the Lerma Valley (Salta, Argentina)," *Renew Energy*, vol. 99, pp. 57–68, Dec. 2016, doi: 10.1016/j.renene.2016.06.042.

[35] † Christelle Rigollier, O. Bauer, and L. Wald´´, "ON THE CLEAR SKY MODEL OF THE ESRA-EUROPEAN SOLAR RADIATION ATLAS-WITH RESPECT TO THE HELIOSAT METHOD," 2000. [Online]. Available: www.elsevier.com/locate/solener

[36] A. Grover, A. Kapoor, and E. Horvitz, "A deep hybrid model for weather forecasting," in *Proceedings of the ACM SIGKDD International Conference on Knowledge Discovery and Data Mining*, Association for Computing Machinery, Aug. 2015, pp. 379–386. doi: 10.1145/2783258.2783275.

[37] T. Saba, A. Rehman, and J. S. AlGhamdi, "Weather forecasting based on hybrid neural model," *Appl Water Sci*, vol. 7, no. 7, pp. 3869–3874, Nov. 2017, doi: 10.1007/s13201-017-0538-0.

[38] H. Sangrody, M. Sarailoo, N. Zhou, N. Tran, M. Motalleb, and E. Foruzan, "Weather forecasting error in solar energy forecasting," *IET Renewable Power Generation*, vol. 11, no. 10, pp. 1274–1280, Aug. 2017, doi: 10.1049/iet-rpg.2016.1043.

[39] J. Mohan, A. Gupta, Jaypee Institute of Information Technology University, Institute of

Electrical and Electronics Engineers. Uttar Pradesh Section. SP/CS Joint Chapter, and Institute of Electrical and Electronics Engineers, *2019 International Conference on Signal Processing and Communication (ICSC) : 07-09 March 2019, Jaypee Institute of Information Technology, NOIDA*.

[40] A. K. Yadav, H. Malik, and S. S. Chandel, "Selection of most relevant input parameters using WEKA for artificial neural network based solar radiation prediction models," *Renewable and Sustainable Energy Reviews*, vol. 31. Elsevier Ltd, pp. 509–519, 2014. doi: 10.1016/j.rser.2013.12.008.

[41] A. Hasni, A. Sehli, B. Draoui, A. Bassou, and B. Amieur, "Estimating global solar radiation using artificial neural network and climate data in the south-western region of Algeria," in *Energy Procedia*, Elsevier BV, 2012, pp. 531–537. doi: 10.1016/j.egypro.2012.05.064.

[42] D. A. Fadare, "Modelling of solar energy potential in Nigeria using an artificial neural network model," *Appl Energy*, vol. 86, no. 9, pp. 1410–1422, 2009, doi: 10.1016/j.apenergy.2008.12.005.

[43] S. Kumar and T. Kaur, "Development of ANN Based Model for Solar Potential Assessment Using Various Meteorological Parameters," in *Energy Procedia*, Elsevier Ltd, Dec. 2016, pp. 587–592. doi: 10.1016/j.egypro.2016.11.227.

[44] † A Sfetsos and A. H. Coonick, "UNIVARIATE AND MULTIVARIATE FORECASTING OF HOURLY SOLAR RADIATION WITH ARTIFICIAL INTELLIGENCE TECHNIQUES," 2000. [Online]. Available: www.elsevier.com/locate/solener

[45] L. Mazorra Aguiar, B. Pereira, M. David, F. Díaz, and P. Lauret, "Use of satellite data to improve solar radiation forecasting with Bayesian Artificial Neural Networks," *Solar Energy*, vol. 122, pp. 1309–1324, Dec. 2015, doi: 10.1016/j.solener.2015.10.041.

[46] M. Yorukoglu and A. N. Celik, "A critical review on the estimation of daily global solar radiation from sunshine duration," *Energy Convers Manag*, vol. 47, no. 15–16, pp. 2441–2450, Sep. 2006, doi: 10.1016/j.enconman.2005.11.002.

[47] K. Abhishek, M. P. Singh, S. Ghosh, and A. Anand, "Weather Forecasting Model using Artificial Neural Network," *Procedia Technology*, vol. 4, pp. 311–318, 2012, doi: 10.1016/j.protcy.2012.05.047.

[48] W. S. Lamberson, M. J. Bodner, J. A. Nelson, and S. A. Sienkiewicz, "The Use of Ensemble Clustering on a Multimodel Ensemble for Medium-Range Forecasting at the Weather Prediction Center," *Weather Forecast*, vol. 38, no. 4, pp. 539–554, Apr. 2023, doi: 10.1175/WAF-D-22-0154.1.

[49] P. Hewage *et al.*, "Temporal convolutional neural (TCN) network for an effective weather forecasting using time-series data from the local weather station," *Soft comput*, vol. 24, no. 21, pp. 16453–16482, Nov. 2020, doi: 10.1007/s00500-020-04954-0.

[50] H. T. C. Pedro and C. F. M. Coimbra, "Nearest-neighbor methodology for prediction of intra-hour global horizontal and direct normal irradiances," *Renew Energy*, vol. 80, pp. 770–782, Aug. 2015, doi: 10.1016/j.renene.2015.02.061.

[51] J. Huang and M. Perry, "A semi-empirical approach using gradient boosting and k-nearest neighbors regression for GEFCom2014 probabilistic solar power forecasting," *Int J Forecast*, vol. 32, no. 3, pp. 1081–1086, 2016, doi: 10.1016/j.ijforecast.2015.11.002.

[52] M. Ratshilengo, C. Sigauke, and A. Bere, "Short-term solar power forecasting using genetic algorithms: An application using south african data," *Applied Sciences (Switzerland)*, vol. 11, no. 9, May 2021, doi: 10.3390/app11094214.

[53] F. Wang, Z. Zhen, B. Wang, and Z. Mi, "Comparative study on KNN and SVM based weather

classification models for day ahead short term solar PV power forecasting," *Applied Sciences (Switzerland)*, vol. 8, no. 1, Dec. 2017, doi: 10.3390/app8010028.

[54] S. C. Lim, J. H. Huh, S. H. Hong, C. Y. Park, and J. C. Kim, "Solar Power Forecasting Using CNN-LSTM Hybrid Model," *Energies (Basel)*, vol. 15, no. 21, Nov. 2022, doi: 10.3390/en15218233.

[55] Institute of Electrical and Electronics Engineers and M. IEEE Systems, *2016 IEEE International Conference on Systems, Man, and Cybernetics (SMC) : 9-12 Oct. 2016*.

[56] N. Elizabeth Michael, M. Mishra, S. Hasan, and A. Al-Durra, "Short-Term Solar Power Predicting Model Based on Multi-Step CNN Stacked LSTM Technique," *Energies (Basel)*, vol. 15, no. 6, Mar. 2022, doi: 10.3390/en15062150.

[57] C. H. Liu, J. C. Gu, and M. T. Yang, "A Simplified LSTM Neural Networks for One Day-Ahead Solar Power Forecasting," *IEEE Access*, vol. 9, pp. 17174–17195, 2021, doi: 10.1109/ACCESS.2021.3053638.

[58] A. Yu. Dolganov and Institute of Electrical and Electronics Engineers, *SIBIRCON : International Multi-Conference on Engineering, Computer and Information Sciences (2019 SIBIRCON) : conference proceedings : 21-22 Oct 2019 Academpark, 23-24 Oct 2019 Dom Uchyonykh, Tusur, 25-27 Oct 2019 Ural Hi-Tech Park*.

[59] L. Benali, G. Notton, A. Fouilloy, C. Voyant, and R. Dizene, "Solar radiation forecasting using artificial neural network and random forest methods: Application to normal beam, horizontal diffuse and global components," *Renew Energy*, vol. 132, pp. 871–884, Mar. 2019, doi: 10.1016/j.renene.2018.08.044.

[60] IEEE-Tunisia Section, IEEE Power & Energy Society. Tunisia Chapter, and Institute of Electrical and Electronics Engineers, *GECS'2017 : International Conference on Green Energy & Conversion Systems : 23-25 March, 2017, Hammamet, Tunisia*.

[61] D. Liu and K. Sun, "Random forest solar power forecast based on classification optimization," *Energy*, vol. 187, Nov. 2019, doi: 10.1016/j.energy.2019.115940.

[62] X. Li *et al.*, "Probabilistic solar irradiance forecasting based on XGBoost," *Energy Reports*, vol. 8, pp. 1087–1095, Aug. 2022, doi: 10.1016/j.egyr.2022.02.251.

[63] J. Ye, B. Zhao, and H. Deng, "Photovoltaic Power Prediction Model Using Pre-train and Fine-tune Paradigm Based on LightGBM and XGBoost," in *Procedia Computer Science*, Elsevier B.V., 2023, pp. 407–412. doi: 10.1016/j.procs.2023.09.056.

[64] D. J. Bae, B. S. Kwon, and K. Bin Song, "XGboost-based day-ahead load forecasting algorithm considering behind-the-meter solar PV generation," *Energies (Basel)*, vol. 15, no. 1, Jan. 2022, doi: 10.3390/en15010128.

[65] C. N. Obiora, A. Ali, and A. N. Hasan, "Implementing extreme gradient boosting (xgboost) algorithm in predicting solar irradiance," in *2021 IEEE PES/IAS PowerAfrica, PowerAfrica 2021*, Institute of Electrical and Electronics Engineers Inc., Aug. 2021. doi: 10.1109/PowerAfrica52236.2021.9543159.

[66] P. Probst, M. Wright, and A.-L. Boulesteix, "Hyperparameters and Tuning Strategies for Random Forest," Apr. 2018, doi: 10.1002/widm.1301.

[67] D. Markovics and M. J. Mayer, "Comparison of machine learning methods for photovoltaic power forecasting based on numerical weather prediction," *Renewable and Sustainable Energy Reviews*, vol. 161, Jun. 2022, doi: 10.1016/j.rser.2022.112364.

[68] L. Torre-Tojal, A. Bastarrika, A. Boyano, J. M. Lopez-Guede, and M. Graña, "Above-ground biomass estimation from LiDAR data using random forest algorithms," *J Comput Sci*, vol. 58, Feb. 2022, doi: 10.1016/j.jocs.2021.101517.

[69] L. Wang, "Research and Implementation of Machine Learning Classifier Based on KNN," in *IOP Conference Series: Materials Science and Engineering*, IOP Publishing Ltd, Dec. 2019.

[70] X. Luo, D. Li, Y. Yang, and S. Zhang, "Spatiotemporal traffic flow prediction with KNN and LSTM," *J Adv Transp*, vol. 2019, 2019, doi: 10.1155/2019/4145353.

[71] S. Zhang, X. Li, M. Zong, X. Zhu, and D. Cheng, "Learning k for kNN Classification," *ACM Trans Intell Syst Technol*, vol. 8, no. 3, Jan. 2017, doi: 10.1145/2990508.

[72] S. Zaheer *et al.*, "A Multi Parameter Forecasting for Stock Time Series Data Using LSTM and Deep Learning Model," *Mathematics*, vol. 11, no. 3, Feb. 2023, doi: 10.3390/math11030590.